\documentclass[12pt,preprint]{aastex}
\usepackage{epsfig}
\usepackage{rotating}
\usepackage{lscape}

\shorttitle{Water maser emission in submm galaxies}
\shortauthors{Edmonds et al.}
\slugcomment{AJ draft: Oct 21, 2008}

\def\ga{\mathrel{\raise0.35ex\hbox{$\scriptstyle >$}\kern-0.6em
\lower0.40ex\hbox{{$\scriptstyle \sim$}}}}
\def\la{\mathrel{\raise0.35ex\hbox{$\scriptstyle <$}\kern-0.6em
\lower0.40ex\hbox{{$\scriptstyle \sim$}}}}

\def\hcn{HCN~{\it J}=1-0 }
\def\hij{high-{\it J}~}
\def\loj{low-{\it J}~}

\begin{document}

\title{An EVLA search for water megamaser emission in the submm galaxy
 SMM~J16359+6612 at z=2.5}

\author{
R.~Edmonds,$^{2}$ J.~Wagg,$^{2}$  E.~Momjian,$^{2}$ C.L.~Carilli,$^{2}$  D.J.~Wilner,$^{3}$
 E.M.L.~Humphreys,$^{3}$ K.M.~Menten,$^{4}$ and D.H.~Hughes$^{5}$}

\affil{$^{2}$National Radio Astronomy Observatory, PO Box O, Socorro,
  NM, USA 87801}\email{jwagg@aoc.nrao.edu}

\affil{$^{3}$Harvard-Smithsonian Center for Astrophysics,
             Cambridge, MA, 02138}

\affil{$^{4}$Max-Planck Institut f{\"u}r Radioastronomie, Auf dem H{\"u}gel 69, D-53121 Bonn, Germany}

\affil{$^{5}$Instituto Nacional de Astrof\'isica, \'Optica y Electr\'onica 
             (INAOE), Aptdo. Postal 51 y 216, Puebla, Mexico}

\begin{abstract}
 Using the Expanded Very Large Array, we have conducted a search for 22.2~GHz H$_2$O megamaser emission in 
the strongly lensed submm galaxy, SMM~J16359+6612 at z=2.517. This object is lensed into three components, 
 and after a correction for magnification is applied to its submm-wavelength flux density, it 
 is typical of the bulk of the high-redshift, submm galaxy population responsible for the 
850~$\mu$m extragalactic background ($S_{850\mu m} \sim $~1~mJy). We do not detect any H$_2$O megamaser 
 emission, but the lensing allows us to place an interesting constraint on the luminosity of any 
 megamasers  present, $L_{H_2 O} < $5305~$L_{\odot}$ for an assumed linewidth of
 80~km~s$^{-1}$. Because the far-infrared luminosity in submm galaxies is mainly powered by 
 star formation, and very luminous H$_2$O megamasers are more commonly associated with 
 quasar activity, it could be that blind searches for H$_2$O
 megamasers will not be an effective means of determining redshifts
 for less luminous members of the submm galaxy
 population. 
\end{abstract}


\section{Introduction}
\label{sec:intro}

Single-dish submm/mm-wavelength surveys have revealed a population of massive, dust-enriched,
 far-infrared (FIR) luminous star-forming galaxies in the early Universe
(Smail, Ivison \& Blain 1997; Hughes et al.\ 1998; Barger et al.\ 1998; Bertoldi et al.\ 2000).
Since their discovery, redshift determination for individual members of this submm galaxy 
(SMG) population has
 been hindered by the lack of sufficiently accurate astrometry needed to identify multi-wavelength 
counterparts, limited by the coarse angular resolution of single-dish submm/mm telescopes
($\sim$10-20$''$), and the faintness of SMGs at optical wavelengths. For the brightest members of 
the SMG population, deep radio interferometry imaging  
has provided a powerful solution to the former problem of identifying counterparts 
(e.g. Ivison et al.\ 1998, 2000, 2002, 2007; Smail et al.\ 2000; Webb et al.\ 2003; Clements et al.\ 2004;
 Borys et al.\ 2004;  Dannerbauer et al.\ 2004), resulting in the ability to measure spectroscopic 
redshifts for the proposed optical counterparts to the members of the SMG 
 population. Such surveys find a median
spectroscopic redshift for bright SMGs
($S_{850\mu m} > $5~mJy) of $z \sim 2.2$ (Chapman et al.\ 2003, 2005). However, 
 between 65--80\% of bright SMGs have secure radio counterparts (e.g. Ivison et al.\ 2005, 2007), 
so a significant fraction may lie at $z > 3.5$, given the radio wavelength
 selection effects. The uncertainty in the redshift distribution
of the fainter SMG population ($<$5~mJy at 850$\mu$m) makes it difficult to
complete the luminosity function for SMGs and hence difficult to accurately
constrain the obscured star-formation rate of the Universe at the highest
redshifts.

Alternative methods of redshift determination for SMGs have been explored, including photometric
 redshift techniques which rely on the measured continuum flux densities
 at submm-to-radio wavelengths
 (Hughes et al.\ 1998; Carilli \& Yun 1999, 2000; Aretxaga et al.\ 2003, 2005). These
 techniques have proven to be accurate to $\delta z \pm 0.3$ (rms) for SMGs at $z \la 3.5$,
when three or more FIR-to-radio flux densities are measured (Aretxaga
 et al.\ 2005)\footnote{It has been argued that this same level of uncertainty can
 be achieved if one assumes 
the median redshift of the bright SMG sample for individual objects (Chapman et al.\ 2005).}.
  For those SMGs with well constrained spectral energy distributions, the predictions of 
such techniques have helped to guide the choice of tunings for broadband
cm-wavelength searches for redshifted CO line emission (Wagg et al.\ 2007), which should be
 an effective means of measuring SMG redshifts given the intensity of molecular CO line
 emission in these objects (e.g. Neri et al.\ 2003; Greve et al.\
 2005). The first cm-wavelength search for redshifted \loj CO line
 emission (\textit{J}=2-1/1-0) 
 in a SMG with no optical spectroscopic redshift was unsuccessful
 (Wagg et al.\ 2007), while
 a great deal of uncertainty remains in the expected intensity in the \loj CO lines. 
 Given the expected high densities of the molecular gas in the
 interstellar media of these SMGs (Weiss et al.\ 2005), it is possible that mm-wavelength
 searches for \hij CO line emission will yield more positive results. Such 
 observations will soon be possible with the ultra-wide bandwidth ``Redshift Receiver''
 on the Large Millimeter Telescope (LMT; P{\'e}rez-Grovas et al.\ 2006).

Broadband spectral line observations of SMGs at lower frequencies ($\nu_{obs} < $10~GHz), may
 yield redshifts through the detection of OH or H$_{2}$O maser emission
(Townsend et al.\ 2001; Ivison 2006). Such observations have an advantage over higher frequency
 CO line searches in that the bandwidth needed to cover the same redshift interval decreases with
 decreasing frequency. This means that the broadband capabilities of the NRAO Green Bank Telescope
(GBT\footnote{The NRAO is operated by Associated Universities Inc., under a cooperative
 agreement with the National Science Foundation.}), or the Expanded Very Large Array (EVLA),
could potentially be exploited to measure redshifts for SMGs via OH or H$_2$O maser emission
 lines. The luminosity in the 22.2351~GHz H$_2$O  maser line has been found to depend weakly
 on FIR luminosity in nearby galaxies (Castangia et al.\ 2008), while a stronger dependence
 of the OH maser line luminosity on FIR luminosity has been demonstrated
(Darling \& Giovanelli 2002). However, given that the OH maser lines emit at rest frequencies
 of 1665.4018 and 1667.3590~MHz, for objects at $z > 2$, these lines are redshifted to the
$\nu_{obs} \la 600$~MHz band, where the presence of radio frequency interference (RFI) 
 for many existing facilities, may make it difficult to identify the OH lines in
 objects whose redshifts are unknown. For these same redshifts, the H$_2$O  maser line is 
redshifted to frequencies in the
 5--8~GHz range, where RFI is less of a problem. It should then be more effective to conduct
 blind searches for redshifted H$_2$O  maser emission when attempting to measure redshifts for
SMGs. Before such a technique can be applied, it must first be demonstrated that  H$_2$O  masers
 exist in SMGs with known redshifts. To date, the highest redshift detection of
 an H$_2$O  maser is in a type 2 quasar at $z = 0.66$ (Barvainis \& Antonucci 2005). Ivison (2006)
 conducted a search for both OH and H$_2$O megamaser emission in the
 strongly lensed, ultraluminous 
 infrared quasar, APM~08279+5255 at z=3.91, however no line emission is detected.

Here, we conduct a search for H$_2$O  maser emission in the lensed SMG, SMM\,J16359+6612 at
$z = 2.517$ (Kneib et al.\ 2004, 2005). Given the large total magnification factor inferred by the
 gravitational lensing model ($m \sim 45$), this object has an intrinsic flux density typical
 of the bulk of the SMG population responsible for the 850~$\mu$m background radiation
($S_{850\mu m} \sim$1~mJy). This object is strongly gravitationally lensed,
 and is the only SMG to have been detected in \hcn line emission (Gao et al.\ 2007),  
 implying the presence of dense gas. 
SMM\,J16359+6612 presents a good candidate in which to conduct a search for H$_2$O  
maser emission, and one of only two strongly lensed SMGs for which
such a search is possible. Our overall aim is to determine if searches
for this line can be used to 
 obtain blind redshifts for SMGs in future surveys, and for this pilot
 study we have chose to focus on less luminous SMGs typical of the
 submm extragalactic background. Throughout this
 work, we assume H$_{0}$=71\,km\,s$^{-1}$, $\Omega$$_{M}$=0.27, and
$\Omega$$_{\Lambda}$=0.73 (Spergel et al.\ 2007).

\section{Observations and data reduction}
\label{sec:obs}

We observed SMM\,J16359+6612 with the Very Large Array (VLA) on 2008 June 4 in its compact 
 DnC configuration. The 
 details of these observations are given in Table~1. At the source redshift of $z = 2.517$
(determined from multiple CO emission lines; Weiss et al.\ 2005), the 22.23508~GHz H$_{2}$O 
maser line is redshifted to 6.322~GHz, which is now accessible with the new C-band receivers 
 available through the Expanded Very Large Array (EVLA) project. The observations were
 configured to have a 391\,kHz channel width, equivalent to 18.5\,km~s$^{-1}$ velocity resolution 
over a bandwidth of 12.5~MHz (591\,km~s$^{-1}$).
 In total, $\sim$9\,hours of on-source integration time was obtained with
 the 15 EVLA antennas available, leading to a synthesized beamsize of
 $\sim$10$''$. Due to the current C-band receiver temperatures at
 non-standard frequencies, and a source of unknown noise associated with using
 12.5~MHz of bandwidth with the VLA correlator, our final sensitivities
 are $\sim$5$\times$ worse than what will be achievable using
 the full array of EVLA antennas with the new, WIDAR correlator.

Data reduction was performed using the Astronomical Image Processing
System (AIPS) of NRAO. 
Some ``flagging'' was required, which resulted in 2 of the 15 EVLA
antennas being removed entirely from the analysis. 
Flux, bandpass, and phase calibrations were derived from observations of 3C286 and 1642+689. After 
applying the calibration solutions to the target data, uniform weighted dirty stokes I 
cubes were produced.

\section{Results}
\label{sec:results}

\begin{center}
\begin{figure}[p]
\includegraphics[scale=0.75]{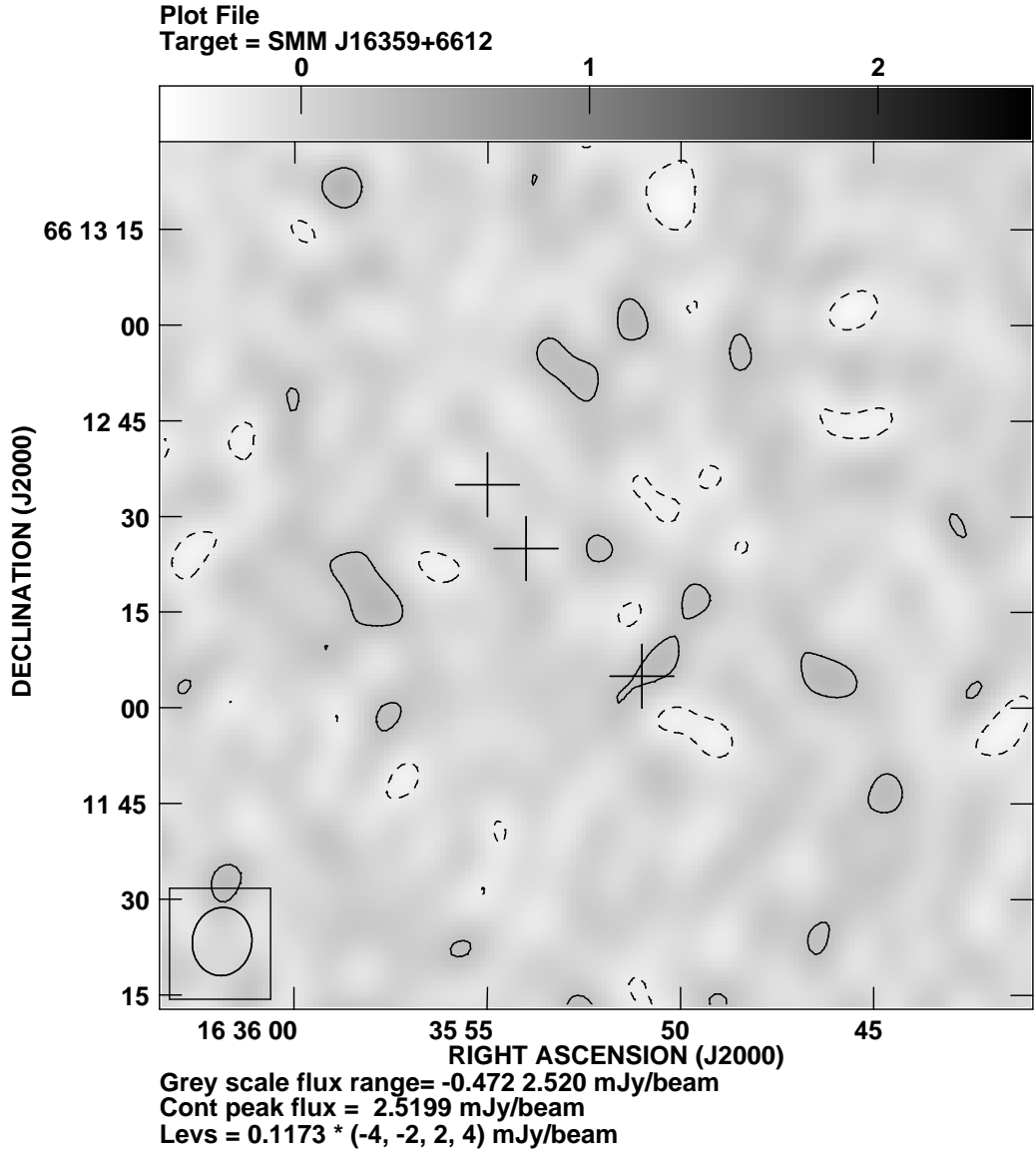}
\caption{Contour map of the 6.3~GHz continuum emission in the SMM~J16359+6612 field. The positions of the  
three lensed components detected in submm continuum and CO line emission are indicated by crosses.
 From left-to-right are components A, B and C, as discussed in the
 text. Contour levels are -2 and 2$\times\sigma$, where $\sigma = 117$~$\mu$Jy~beam$^{-1}$.}
\label{fig1}
\end{figure}
\end{center}

\begin{center}
\begin{figure}[p]
\includegraphics[scale=0.75]{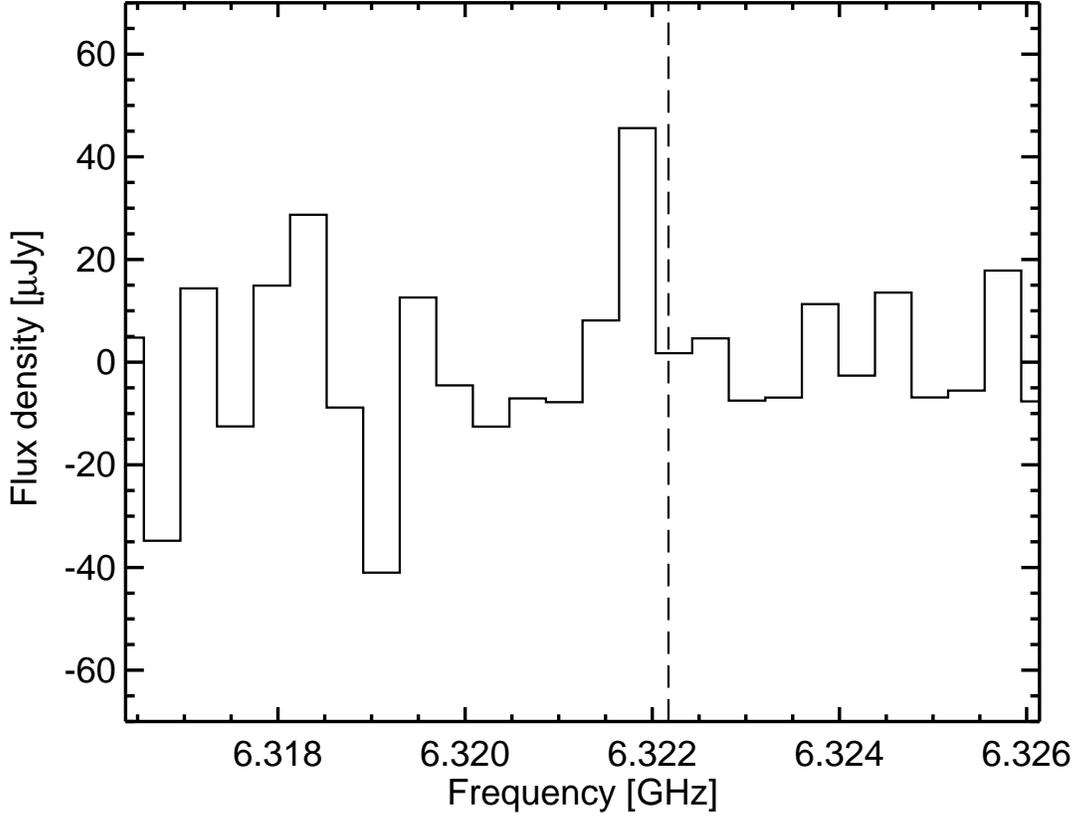}
\caption{10~MHz of the average C-band spectrum of the 3 lensed components of  
SMM~J16359+6612, after correcting each for the appropriate
magnification factor. The rms of this spectrum is $\sim$16~$\mu$Jy~beam$^{-1}$ per 
391~kHz channel. The vertical dashed line shows the emission
frequency expected if the redshift were the same as that of the \hij CO lines.}
\label{fig2}
\end{figure}
\end{center}

The submm/mm emission from SMM\,J16359+6612 is lensed into 3 components, where the observed flux densities 
 of components A, B and C have been amplified by factors of 14, 22, and 9, respectively (Kneib et al.\ 2004).
 No 6.3~GHz continuum emission is detected from any of the 3
 components in the continuum image above an apparent 
rms of $\sim$117~$\mu$Jy/beam (Fig.\,\ref{fig1}). These limits on the 6.3~GHz continuum 
 emission are consistent with the value of 175~$\mu$Jy for the peak 6.3~GHz flux density of the brightest 
component, predicted from the 1.4 and 8.2~GHz detections of SMM\,J16359+6612
with the VLA and the  
 Westerbork Synthesis Radio Telescope (WSRT) by Garrett et al.\ (2005).  
Spectra were extracted at the positions of 
each of the lensed components determined from the centroids of the \hij CO line emission. 
Each spectrum was corrected for lensing amplification. From the
 three spectra, an average spectrum was created using the
 ``de-magnified'' noise of each in the weighting (Fig.\,\ref{fig2}). Although our channel rms is
 $\sim$520~$\mu$Jy/beam, by first de-magnifying and then averaging the
 spectra extracted at the positions of the three components, we are
 able to reach a sensitivity equivalent 
 to $\sim$16~$\mu$Jy/beam per 391~kHz channel.

Inspection of the individual spectral line data cubes did not reveal
any significant detection of H$_2$O megamaser 
emission in SMM\,J16359+6612, nor is any emission detected in the average spectra of the 3 components 
(Fig.\,\ref{fig2}). 
Although we do not detect redshifted H$_2$O maser emission in SMM\,J16359+6612, our sensitivity is 
 sufficient to derive strong constraints on the luminosity of any H$_2$O megamaser emission present. 
 Typical linewidths of H$_2$O maser emission lines in FIR
  bright galaxies are in the range 1--130~km~s$^{-1}$, with one
  example of a $\sim$260~km~s$^{-1}$ line in Arp299 (Henkel et
  al.\ 2005). As such, we assume intermediate
linewidths of 20, 40, 60 and 80~km~s$^{-1}$, in order to calculate our H$_2$O line 
luminosity limits following the definition given in Solomon \& Vanden Bout (2005). Before
 correcting the noise of an individual component for lensing
 magnification, the 3-$\sigma$ line luminosity 
limits are, $L_{H_2 O} < $~(8.7, 12.2, 15.0, and 17.3)$\times 10^{4}$~$L_{\odot}$, for
 the assumed linewidths. Much stronger constraints are obtained when
 we correct for lensing and calculate these luminosity limits
 for the average spectrum: $L_{H_2 O} < $~2652, 3751, 4594, and
 5305~$L_{\odot}$.

\section{Discussion}
\label{sec:discuss}

We can estimate whether our limits to the H$_2$O megamaser line
 luminosity in SMM~J16359+6612 are 
  consistent with the value expected from the FIR luminosity and the 
 $L_{FIR} - L_{H_2 O}$ luminosity correlation suggested by Castangia
 et al.\ (2008) for nearby galaxies. We note that this correlation is weak, and a wide
 range of H$_2$O line luminosities appear to be valid. The unlensed 
 850~$\mu$m flux density of each image of SMM~J16359+6612 is $\sim$0.8~mJy (Kneib et al\ 2004), 
 which can be converted 
 to a FIR luminosity, $L_{FIR} \sim $1.5$\times$10$^{12}$~$L_{\odot}$
 for an assumed dust temperature,
 $T_d = 40$~K. From the $L_{FIR} - L_{H_2 O}$ luminosity relation for
 Galactic star-forming regions 
(Genzel \& Downes 1979; Jaffe et al.\ 1981), we should expect an H$_2$O megamaser line
 luminosity, $L_{H_2 O} \sim $1500~$L_{\odot}$. This prediction is
 broadly consistent with the limits we obtain here for the narrowest
 linewidths assumed. For comparison, the luminosity in the H$_2$O megamaser line detected 
in a type 2 quasar at $z=0.66$ by Barvainis \& Antonucci (2005) is 2.3$\times$10$^4$~$L_{\odot}$.

We propose possible explanations for our non-detection of H$_2$O megamaser
 emission in SMM~J16359+6612. 
Although it is unlikely, it could be that the limited velocity covered by our choice of bandwidth
 ($\sim$574\,km\,s$^{-1}$) is too small, as the maser emission may be
 at a significantly different redshift from that of the CO 
 emission lines (e.g. Wilner et al.\ 1999). 
However, the more probable reason for our non-detection is that
 the molecular medium within SMM\,J16359+6112 is not conducive to the
 formation of megamaser emission. Neufeld et al.\ (1994) argue that
 luminous water maser emission can be excited by X-ray photons from an
 active galactic nuclei (AGN). Strong H$_{2}$O maser emission would
 arise in gas with temperatures of $\sim$400-1000\,K and pressures of
 P/$k$=$\sim$10$^{11}$\,cm$^{-3}$\,K, irradiated by intense X-ray emission. 
 Such AGN powered circumnulear H$_2$O megamasers can attain much larger
line luminosities than the interstellar masers found in starburst galaxies,
which are consequently termed ``kilomasers''.  
  X-ray studies of SMGs confirm that while $\sim$28--50\% do 
harbour an AGN, such AGN activity may only contribute to a small
 fraction of their enormous FIR luminosities (e.g. Alexander et al.\
 2005).
Given the rarity, and lower luminosities of AGN in SMGs, it is likely that the
 dense molecular gas in SMM~J16359+6612 inferred from the HCN line
 emission is most likely associated with high-mass star-formation.  
Extragalactic regions of star-formation may give rise to water maser emission,   
 however this emission is generally weaker than that observed in
 galaxies with luminous AGN. Such is the case with the nearby galaxy, NGC\,520, whose
 kilomaser line luminosity is lower than expected from its FIR luminosity (Castangia et al.\ 2008). 
 A third possibility for our non-detection of H$_2$O megamaser emission is
 that the emission is variable. Argon et al.\ (2007) found that the
 H$_2$O maser emission in the nearby galaxy, NGC~4258, varied by a
 factor of $\sim$12 over a 3 year period.

 This non-detection of H$_2$O megamaser emission in
SMM~J16359+6612 has consequences for the use of this line 
 as a redshift indicator in future studies of less luminous
   SMGs typical of the submm extragalactic background. Given the strong gravitational 
 lensing amplification of this object, making it one of the brightest 
 of the currently detected SMGs, J16359 
 should be one of the most likely SMGs to exhibit detectable 
 H$_2$O megamaser emission.
 It has recently been suggested that the 
 H$_{2}$O maser detection rate is weakly correlated with a galaxy's FIR luminosity 
(Castangia et al.\ 2008). 
 If this is the case, then it may be that
 similarly bright, unlensed SMGs may be more likely to contain  H$_2$O
 megamasers. However, mounting evidence suggests that the 
brightest ($S_{850\mu m} \ga $15~mJy), and subsequently most luminous
SMGs may be at the highest redshifts 
 (e.g. Younger et al.\ 2007; Wang et al.\ 2007), and
 so cosmological dimming of the line intensity would necesitate
 large integration times (more than 100~hours), even with the full
 EVLA. Such searches would be best carried out with the future Square
 Kilometer Array (SKA). 

\section{Summary}
\label{sec:sum}

We have conducted the first search for 22~GHz water megamaser emission in a
high-redshift SMG, by observing 
the strongly lensed, SMM\,J16359+6612 at z=2.517, with the currently
available EVLA antennas. 
No emission is detected, resulting in a 3-$\sigma$ luminosity limit of
$<$5305\,L$_{\odot}$ for an assumed linewidth of 80~km~s$^{-1}$.
 With an optimally working, completed EVLA we would have achieved
 $\sim$5$\times$ higher sensitivity, reaching down into the kilomaser
 luminosity range. Our non-detection suggests that luminous H$_{2}$O megamasers
  may not be present to provide reliable redshift estimates for the less
   luminous members of the SMG population, however future surveys of 
  lensed SMGs with secure CO line redshifts are needed
 to strengthen this conclusion.

\section{Acknowledgments}

We thank the NRAO staff involved in
the EVLA project for making these observations possible. 
JW and CC are grateful for support from the Max-Planck Society and the
Alexander von Humboldt Foundation. We thank the anonymous referee for
helpful suggestions on the original manuscript.

\begin{deluxetable}{lc}
\tablecaption{H$_2$O observations of SMM J16359+6612.
\label{table1}}
\tablewidth{0pt}
\tablehead{\colhead{} & \colhead{}}
\startdata
Frequency:& 6.322~GHz  \\
Pointing Center (J2000): & 16$^h$35$^m$44$^s$.15, +66$^o$12$^m$24$^s$  \\
Synthesized beam: & $10\farcs72 \times 9\farcs31$, P.A. -7.4$^{\circ}$ \\
Channel spacing: & 391~kHz\\
line rms (391~kHz channel):   & 520~$\mu$Jy~beam$^{-1}$  \\
continuum rms: & 117~$\mu$Jy~beam$^{-1}$  \\
\enddata            
\end{deluxetable}

\end{document}